\documentclass[amsmath,amssymb,prl,hyperlink,twocolumn]{revtex4}

\usepackage{graphicx}
\usepackage{soul}
\usepackage[colorlinks=true,citecolor=blue,linkcolor=magenta]{hyperref}
\usepackage[usenames]{color}
\usepackage{amsfonts}

\usepackage{times}

\begin{document}


\title{Deterministic and Robust Generation of Single Photons On a Chip with 99.5\% Indistinguishability \\ Using Rapid Adiabatic Passage}

\author{Yu-Jia Wei$^{1}$, Yu-Ming He$^{1}$, Ming-Cheng Chen$^{1}$, Yi-Nan Hu$^1$, Yu He$^{1}$, Dian Wu$^1$, Christian Schneider$^2$,  \\ Martin Kamp$^2$,  Sven H\"{o}fling$^{2,1,3}$, Chao-Yang Lu$^{1}$, Jian-Wei Pan$^1$\vspace{0.2cm}}

\affiliation{$^1$ Hefei National Laboratory for Physical Sciences at Microscale and Department of Modern Physics, University of Science and Technology of China, Hefei, Anhui 230026, China}
\affiliation{$^2$ Technische Physik, Physikalisches Instit\"{a}t and Wilhelm Conrad R\"{o}ntgen-Center for Complex Material Systems, Universitat W\"{u}rzburg, Am Hubland, D-97074 W\"{u}zburg, Germany}
\affiliation{$^3$ SUPA, School of Physics and Astronomy, University of St. Andrews, St. Andrews KY16 9SS, United Kingdom}

\date{\vspace{0.1cm}\today}

\begin{abstract}
We demonstrate deterministic and robust generation of pulsed resonance fluorescence single photons from a single InGaAs quantum dot using the method of rapid adiabatic passage. Comparative study is performed with transform-limited, negatively chirped and positively chirped pulses, identifying the last one to be the most robust against fluctuation of driving strength. The generated single photons are background free, have a vanishing two-photon emission probability of 0.3\% and a raw (corrected) two-photon Hong-Ou-Mandel interference visibility of 97.9\% (99.5\%), reaching a precision that places single photons at the threshold for fault-tolerant surface-code quantum computing. The single-photon source can be readily scaled up to multi-photon entanglement and used for quantum metrology, boson sampling and linear optical quantum computing.
\end{abstract}


\maketitle

Photons offer an appealing platform for quantum information processing because of their fast transmission, low decoherence, and the ease of implementing precise single-qubit operations \cite{Kok-Obrien, pan-RMP}. In the past two decades, a number of seminal experiments in quantum information, such as quantum teleportation \cite{teleportation}, entanglement purification \cite{purification}, quantum metrology \cite{metrology}, quantum computation \cite{computing} and boson sampling \cite{bosonsampling}, have been performed using up to eight photons \cite{eightphoton} produced by parametric down conversion \cite{Kwiat}. Yet, scaling up from these small-scale demonstrations to more, practically useful, photons appears challenging due to the probabilistic generation and high-order emission of photon pairs in parametric down conversion \cite{pan-RMP}.

To overcome these shortcomings, an increasing effort has turned to single quantum emitters \cite{single-emitters}. Self-assembled quantum dots (QDs) on scalable semiconductor chips are promising single-photon emitters with high quantum efficiency \cite{QD-emitters}. Further, QDs can be embedded in monolithic nanocavities to enhance light-matter interaction and photon extraction \cite{nanocavity}. Since the first observation of photon antibunching from QDs \cite{michler}, numerous experiments have demonstrated single-photon emission \cite{QD-emitters, nanocavity, Santori-Nature2002, RF, YMHe}.

Scalable photonic quantum technologies place stringent demands on the single-photon sources. One of the key requirements is deterministic single-photon generation, that is, the source should emit one---and only one---photon upon each pulsed excitation. To this end, ultrafast laser pulses have been used to resonantly excite a QD two-level system to generate on-demand resonance fluorescence (RF) single photons and observe Rabi oscillation (RO) \cite{YMHe}. Under a $\pi$ pulse excitation, the two-level system is deterministically prepared in the excited state, followed by radiative emission of a single photon. However, using this technique, the efficiency of single-photon generation is sensitive to the variation of pulse area, which can be caused by fluctuation of experimental parameters such as the excitation power and dipole moment.

A more robust method of excited state preparation in a two-level system is rapid adiabatic passage (RAP) with frequency-chirped pulses. Unlike RO, adiabatic popular transfer is largely unaffected by the variation in the optical field, interaction time, and atomic dipole moment. Previous experiments have used negatively chirped pulses to demonstrate population transfer in single QDs, where the exciton population is read out by photocurrent \cite{ARP1} or probabilistic photon emission from a tunneling-assisted spectator state \cite{ARP2}. However, access to actual emitted pulsed RF single photons without laser background and time jitter, and their detailed investigations have proven elusive. Additionally, only negatively chirped pulses have been tested, which are in theory less robust than the positively chirped pulses \cite{positive-negative1,positive-negative2,positive-negative3,positive-negative4}.

Meanwhile, another major requirement for future photonic quantum technologies is that the photons should have a high degree of indistinguishability to meet the fault-tolerant threshold for scalable quantum computing. Similarly, other applications such as quantum repeaters \cite{repeater} and boson sampling \cite{bosonsampling} also heavily rely on the photons' indistinguishability \cite{boson-threshold}. To date, Hong-Ou-Mandel interference with single photons on demand from QDs have demonstrated a raw (corrected) visibility up to 91\% (97\%). Yet, a direct observation of photons' indistinguishability unambiguously surpassing the highest threshold of surface-code quantum computing \cite{threshold} remains elusive.

\begin{figure}[thb]
    \centering
        \includegraphics[width=0.38\textwidth]{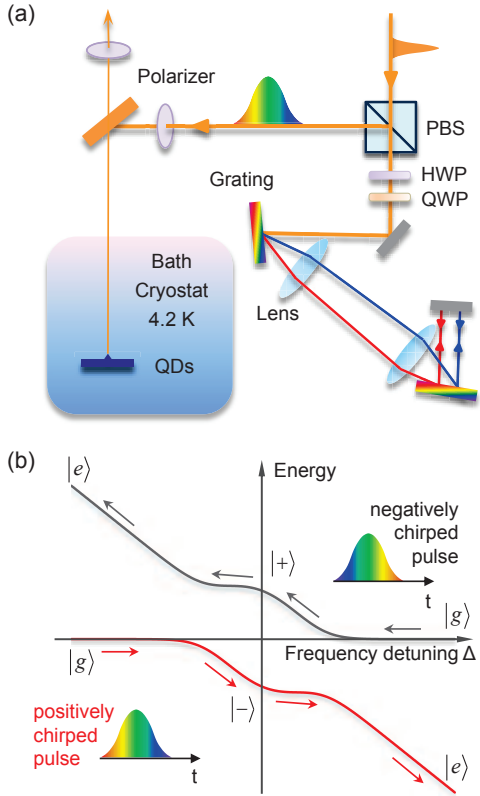}
\caption{
(a) The experimental set-up. The QDs are housed in a stable bath cryostat. The positively-chirped pulses are obtained by stretching the pulse duration of the 3 ps transform-limited laser pulse to 30 ps using two symmetrical gratings sandwiched with a telescope. PBS: polarizing beam splitter. HWP: half wave plate. QWP: quarter wave plate.  (b) The eigenenergies of a two-level system interacting with chirped pulses in the rotation frame. The eigenstates of two-level atom interacting with the resonant laser field are no longer the atomic ground state $|g\rangle$ and excited state $|e\rangle$ but their coherent superposition $|\pm\rangle$ (dressed states). When the frequency of the pump laser is far red (blue) detuned, the states $|+\rangle$ and $|-\rangle$ are reduced to $|e\rangle$ ($|g\rangle$) and $|g\rangle$ ($|e\rangle$). The atom starts from the ground state and evolves adiabatically to the excited state along a dressed state when the frequency of the excitation laser pulse is swept across the excited state, with a rate much slower than the peak Rabi frequency. Using a positively chirped pulse, i.e., the laser sweeps from low to high frequency, the QD evolves along the lower dressed state $|-\rangle$. Using a negatively chirped pulse, the QD follows the upper dressed state $|+\rangle$. The peak Rabi frequency in our experiment is $\sim$100 GHz.
} \label{fig:1}
\end{figure}

Here we demonstrate deterministic and robust generation of single photons from a single QD through RAP with positively chirped pulses. A direct comparison of power-dependent photon generation efficiency is performed with transform-limited, negatively chirped and positively chirped pulses. The generated RF photons show a vanishing two-photon emission probability of 0.3\% and a raw (corrected) two-photon quantum interference visibility of 97.9\% (99.5\%), placing single photons at the fault-tolerant threshold for surface-code quantum computing.

Our experiment is performed on a single negatively charged InGaAs QD cooled to 4.2 K. The QD is embedded in a planar microcavity consisting of 5 upper and 24 lower distributed-Bragg-reflector mirrors. As shown in Fig.$\,$1(a), laser excitation and QD fluorescence collection are operated on a confocal microscope, where a polarizer is placed in the collection arm with its polarization
perpendicular to the excitation light, extinguishing the scattered laser by a factor exceeding $10^6$.

For a comparative study, three different types of laser pulses are used, all resonant with the QD $X^{\mathrm{-1}}$ transition. The first one are transform-limit hyperbolic secant pulses of 3 ps duration from a mode-locked Ti:sapphire laser \cite{YMHe}. The second are negatively chirped pulses of 30 ps duration, generated by a stretcher consisting of two gratings with 1200 lines/mm placed parallel to each other \cite{negativechirped}, similar to the previous experiments \cite{ARP1,ARP2}. The last are positively chirped pulses of 30 ps duration, which are obtained by inserting a telescope into the two symmetrical gratings \cite{positivechirped}, as shown in Fig. 1(a).

\begin{figure*}[tb]
    \centering
        \includegraphics[width=1\textwidth]{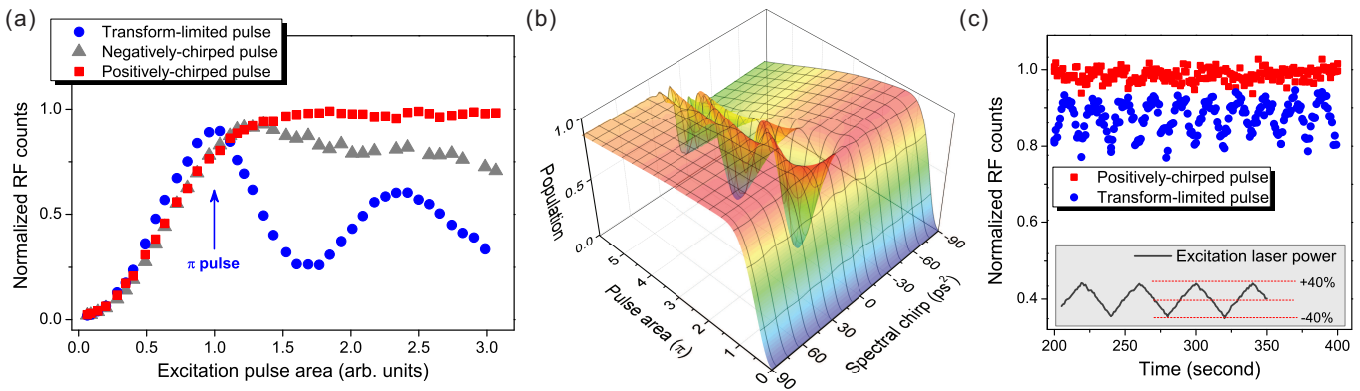}
\caption{
(a) Power-dependent RF count rate for three different excitation methods. (b) Simulation of excited state population as a function of pulse area and spectral chirp. We consider QD-phonon interaction \cite{positive-negative2} in the form $J(\omega)=\alpha\omega^3\textup{exp}[-(\omega/\omega_c)^2]$, where $\alpha$ is the coupling strength and $\omega_c$ is cut-off frequency \cite{EID1}. The temperature, $\alpha$ and $\omega_c$ used in the simulation are 4.2 K, 0.022 $\textup{ps}^{-2}$ and 2 $\textup{ps}^{-1}$, respectively. (c) Time-dependent RF count rates under external modulation of laser power as indicated by the dark line in the inset.
} \label{fig:2}
\end{figure*}

\begin{figure}[tb]
    \centering
        \includegraphics[width=0.38\textwidth]{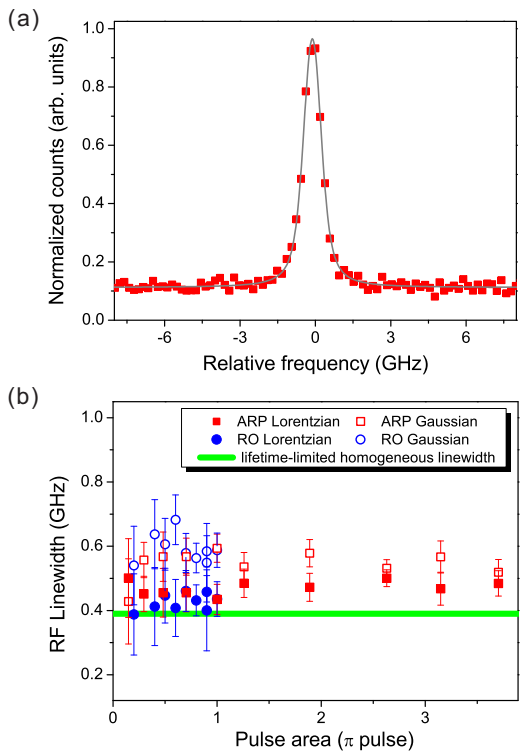}
\caption{(a) A high-resolution spectrum of 1.8$\pi$-pulse excited RF. The solid line is a fit of Voigt profile. (b) The extracted linewidth of the Lorentzian (solid) and Gaussian component (hollow) of the fitted Voigt lines for the RF photons when the QD is driven by positively-chirped (squares) and transform-limited pulses (circles).} \label{fig:3}
\end{figure}

Figure$\,$2(a) displays the detected pulsed RF photon counts as a function of the square root of the excitation laser power. The transform-limited laser pulses yield a sinusoidal oscillation in the emission intensity (blue circles), as a result of the RO in a driven two-level system. The RF photon counts reach the first peak at $\pi$ pulse, indicating that the population is deterministically transferred to excited states at this exact point \cite{YMHe}. The damping at the high power regime could be caused by excitation induced dephasing \cite{EID1,EID2}.

A stark difference is evident when the QD is excited by the chirped pulses. Driven by the 30 ps negatively chirped pulses, the system evolves following the black line in Fig.$\,$1(b), and the RF counts (gray triangles) climbs up to the maximum at the pulse area around 1.5$\pi$. After the maximum, a gradual decrease of the counts is observed, in agreement with Ref.$\,$\cite{ARP2} and our simulation shown in Fig.$\,$2(b). This could be caused by relaxation from the $\left |+\right\rangle$ to the $\left |-\right\rangle$ eigenstate accompanied by acoustic phonon emission that breaks down the RAP \cite{positive-negative1,positive-negative2,positive-negative3,positive-negative4}, which can reduce the efficiency of single-photon generation against excitation power fluctuation.

This problem can be remedied by using positively chirped pulses that drive the RAP along the red line shown in Fig.$\,$1(b). Here, the possible mechanism for disrupting the RAP, caused by the relaxation from the $\left |-\right\rangle$ to the $\left |+\right\rangle$ eigenstate, would require absorption of phonons which is much less likely than emitting them at low temperature. This is in agreement with our simulation shown in Fig.$\,$2(b). The experimental data obtained from the positively chirped pulses are plotted as the red squares in Fig.$\,$2(a). The data points overlap well with those of the negative chirp up to the pulse area of 1.5$\pi$ where the RF counts reach maximum. Importantly, the photon intensity remains steady for increased laser power at least up to 3$\pi$.

Consequently, using positively chirped pulses should avoid the decrease of single-photon generation efficiency that would normally caused by the excitation power fluctuation when using the transform-limited or the negatively chirped pulses. We quantify the robustness of single-photon generation by directly comparing the method of RO and RAP. Figure$\,$2(c) shows the time dependence of the recorded RF photon counts in the presence of artificially engineered excitation power fluctuation, by introducing a 50 mHz triangle modulation with a peak-to-peak amplitude of 80\%, centered around the $\pi$ and $1.9\pi$ pulse area for RO and RAP, respectively. The time trace of the transform-limited pulse excited RF photon counts clearly reproduces the laser modulation with an intensity fluctuation amplitude exceeding $15\%$. In contrast, the time trace generated through RAP using positively chirped pulse keeps stabilized (fluctuation amplitude $\thicksim$$\,$$2.2\%$) over this large power range.

\begin{figure}[tb]
    \centering
        \includegraphics[width=0.41\textwidth]{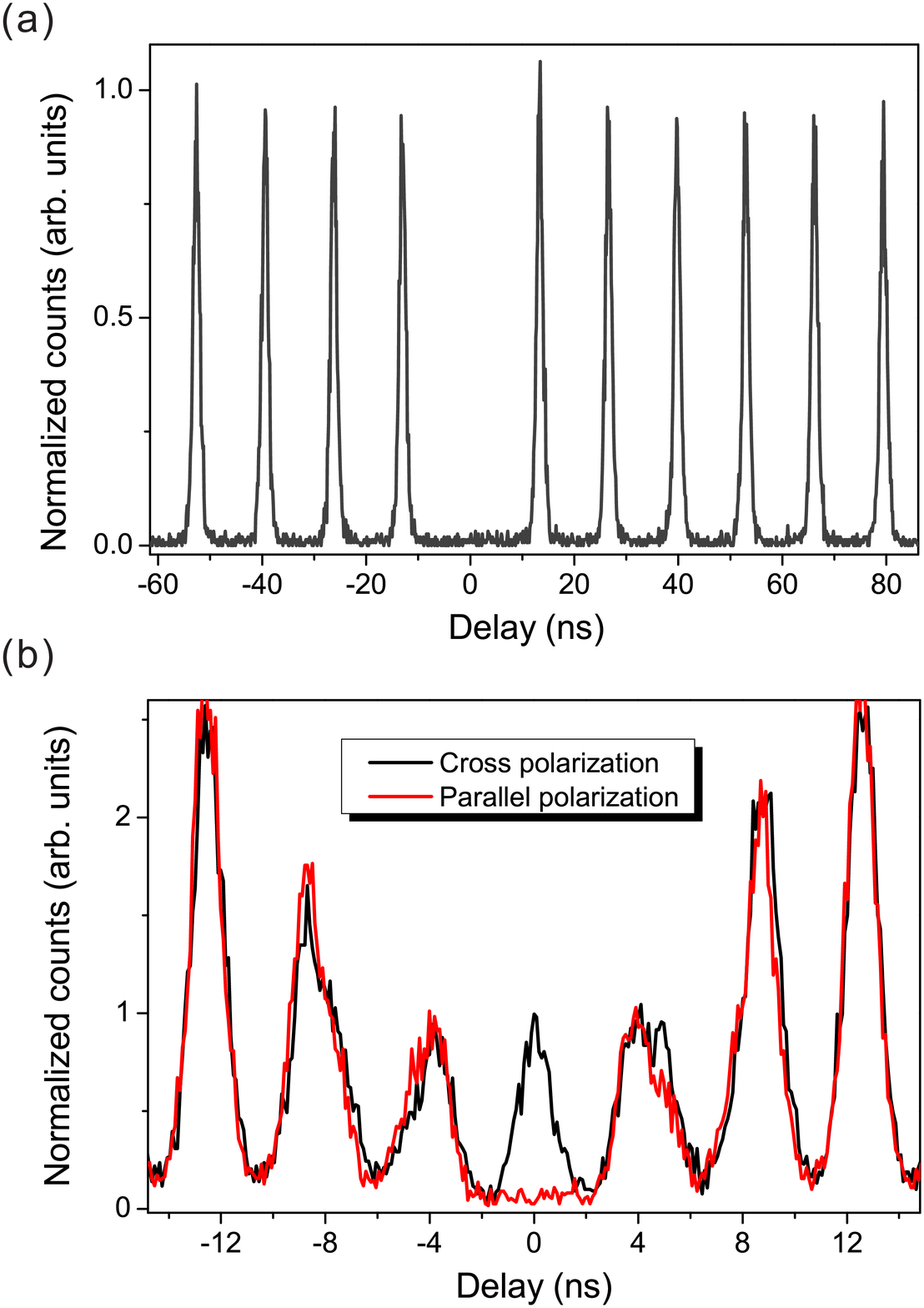}
\caption{(a) Intensity-correlation histogram of the pulsed RF, obtained using a Hanbury Brown and Twiss interferometer. The second-order correlation
 $\textup{g}^2(0)$$\,=\,$$0.003(2)$ is calculated from the integrated photon counts in the zero time delay peak divided by the average of the adjacent six peaks, and its error (0.002)---denoting one standard deviation---is deduced from propagated Poissonian counting statistics of the raw detection events. (b) Non-postselective two-photon Hong-Ou-Mandel interference with two single photons deterministically generated by chirped pulses with a saturated pulse area of 1.6$\pi$. The two consecutive single photons have a delay of 4 ns. The black and red lines are when the two incoming photons are prepared in orthogonal and parallel polarization, respectively. The emitted single photons show an unprecedentedly high indistinguishability, as measured by the raw (corrected) interference visibility of 0.979 (0.995).} \label{fig:4}
\end{figure}

Having demonstrated the robustness of RAP against laser power fluctuation, in what follows we perform a detailed characterization of the generated single photons. High-resolution spectra of the pulsed RF photons are obtained using a scanning Fabry-P$\acute{\textup{e}}$rot interferometer. A typical spectrum is shown in Fig.$\,$3(a) which can be better fitted using a Voigt profile than a Lorentzian lineshape. For a series of excitation powers used in both the RO and RAP methods, the homogeneous (Lorentzian) and inhomogeneous (Gaussian) linewidth are extracted and plotted in Fig.$\,$3(b). Within the $\pi$$-$$3.6\pi$ pulse area range, the RAP homogeneous linewidths remain nearly constant at a level comparable to that at RO $\pi$ pulse. The average homogeneous linewidth is measured to be 0.48(5) GHz, which is close to the lifetime-limited linewidth of 0.39(2) GHz. The Gaussian component in the spectra has an average inhomogeneous linewidth of 0.55(6) GHz, which could be caused by the spectral diffusion due to charge fluctuations in the vicinity of the quantum dot \cite{spectradiffusion}. The undesired spectral diffusion can be mitigated by using microcavity embedded QDs where the Purcell effect could increase the lifetime-limited linewidth to become much larger than the diffusion linewidth. There have also been proposals of using nuclear spin polarization to eliminate this undesired broadening \cite{warburtonnaturephysics}.

To verify that RAP generates truly single photons, we carry out a second-order correlation measurement between the photons emitted from a single QD excited by positively chirped pulse. As displayed in Fig.$\,$4(a), the vanishing of coincidence counts at the zero time delay indicates a near complete suppression of multiphoton emission. The multiphoton probability is calculated to be $\textup{g}^2(0)$$\,=\,$$0.003(2)$ by dividing the total integrated counts within a 3.2 ns window around zero delay by the average counts of the adjacent seven peaks.

Finally, we perform a Hong-Ou-Mandel type interference experiment to analyze the indistinguishability of the single photons. Two identical photons impinging on a 50:50 beam splitter simultaneously will coalesce to the same output spatial mode due to quantum interference. Two positively chirped pulses with 4ns delay, generated with an unbalanced Mach-Zehnder interferometer, excite the QD with a repetition rate of 82 MHz to produce two consecutive single photons. The emitted photons are spectrally filtered with a monochromator and a Fabry-P$\acute{\textup{e}}$rot etalon (with a bandwidth of 1 GHz), and then fed into another unbalanced Mach-Zehnder interferometer, also with a path length difference of 4 ns.

Figure$\,$4(b) displays a time-delayed histogram of the two-photon coincidence count rate. The red and black lines are the data of count rate for two input photons prepared in parallel and cross polarization, respectively. Remarkably, there is a clean suppression of the coincidence counts at zero time delay when the two incoming photons are in the same polarization state. We obtain a raw two-photon interference visibility of 0.979(6), calculated from the integrated counts within a 3.2 ns window around zero time delay for both parallel and cross polarized photons. After correcting with independently calibrated system parameters, including the residual multi-photon probability $\textup{g}^2(0)$$\,=\,$$0.003(2)$ and the first-order interference visibility of 0.995 in the Mach-Zehnder set-up, we obtained the corrected degrees of indistinguishability to be 0.995(7).

We theoretically calculate the process fidelity of two-qubit controlled phase gate affected by the input photon's indistinguishability. The corrected visibility 0.995(7) can in principle yield a gate fidelity of 0.999(3) assuming perfect alignment, beam spitters, and zero dark counts in single-photon detectors. Even with the practical linear optics such as that used in our experiment, a gate fidelity of 0.996(2) can be obtained from the raw visibility of 0.979(6). Both figures of merit surpass the error threshold ($\sim$1\%) for fault-tolerant surface-code quantum computing \cite{threshold}. In this sense, our work demonstrates for the first time that two-photon controlled gate can be operated with a precision sufficient for scalable quantum technologies.

It should be noted that, in addition to the deterministic generation, a truly on-demand single-photon source also requires near-unity photon collection and detection efficiencies. In this experiment, the product of the overall collection efficiency ($\sim$1\%) and detection efficiency ($\sim$20\%) is about 0.2\%, which falls far short of the 67\% efficiency threshold for loss-tolerant optical quantum computing \cite{rudolph}. An essential next step would be to combine the present single-photon source with high-efficiency photon collection \cite{collection} and detection \cite{detection}.

In summary, we have demonstrated robust and deterministic generation of single photons with 99.5\% indistinguishability from a single QD resonantly excited by positively chirped pulses. The sequentially emitted single photons from the same emitter are particularly suitable for the scheme of linear optical quantum computing in a single spatial mode \cite{timebin}. The method can be readily extended to robust preparation of biexciton states in QDs  using two-photon resonance excitation for generation of entangled photons \cite{Liu, michler-twophoton}. Moreover, due to its robustness against inhomogeneous distribution of QD emission wavelength and dipole moment, the same method can be used to excite multiple QDs (on a high-density sample) to achieve deterministic population inversion simultaneously, creating wavelength multiplexed multiple single-photon sources, which could be used for wavelength-division-multiplexing quantum key distribution to enhance the quantum communication capacity. Lastly, the excitation method could also be useful in studying dipole-dipole coupling between two nearby QDs \cite{dipole} and entanglement of two locally separate spins in coupled QDs \cite{entanglegeneration}.

\vspace{0.1cm}
\textit{Acknowledgement}: This work was supported by the National Natural Science Foundation of China, the Chinese Academy of Sciences and the National Fundamental Research Program (under Grant No: 2011CB921300, 2013CB933300), and the State of Bavaria. S.H. acknowledges the CAS visiting professorship. Y.-J.W. and Y.-M.H. contributed equally to this work.

\end{document}